\documentstyle[12pt]{article}
\begin{document}
\def\slash#1{#1 \hskip -0.5em / }
\thispagestyle{empty}
\begin{flushright}
CERN--TH.7086/93 \\ IP-ASTP-30-93 \\ hep-ph/9311327
\end{flushright}
\begin{center}
{\LARGE \bf Subleading Corrections to }    \\
{\LARGE \bf Semileptonic $\Lambda_c$ Decays}
\vspace*{10mm} \\

{\Large \bf Guey--Lin Lin} \vspace*{3mm} \\
{\Large  Institute of Physics,
         Academia Sinica, Taipei}         \\
{\Large  Taiwan 11529, Republic of China } \vspace*{10mm} \\
{\Large \bf Thomas Mannel}    \vspace*{3mm} \\
{\Large CERN -- Theory Division} \\
{\Large CH -- 1211 Geneva 23   } \\
{\Large Switzerland} \vspace*{2cm} \\
\end{center}
\vfill
\begin{center}
{\it Submitted to Phys. Lett B}
\end{center}
\vfill
\begin{center}
{\bf Abstract}

\end{center}
\noindent
The semileptonic decay $\Lambda_c \to \Lambda \ell \nu$ is considered
in the framework of heavy quark effective theory beyond the leading
order in the $1/m_c$ expansion. According to our estimate the
polarization variable will receive only very small corrections such
that $\alpha_{\Lambda_c} \le -0.95$.

\vfill
%
%
%



%
\newpage

\setcounter{page}{1}
\section{Introduction}

The heavy quark limit has become a standard tool for the analysis
of processes involving heavy quarks \cite{IW90,HQET}.
The additional symmetries
present in the infinite mass limit yield useful constraints on
the number of independent form factors parametrizing heavy hadron
weak decays. Furthermore, due to the symmetries one may infer
the normalization of the form factors for transitions among heavy
quarks at certain kinematic points.

Corrections to the limiting behaviour may be calculated systematically
in the framework of heavy quark effective field theory \cite{HQET}
as a power series
expansion in $\alpha_s (m_Q)$ and $\bar\Lambda /m_Q$. Here $m_Q$ is
the mass of the heavy quark and $\bar\Lambda $ is a scale
characterizing
the light degrees of freedom.

Heavy quark symmetries yield also some predictions for the decays of
heavy quarks into light ones \cite{htol}.  However,
in the mesonic case there is no
constraint on the number of independent form factors for weak decays
any more, but heavy flavor symmetry may still be used to relate
semileptonic $B$ decays to the corresponding $D$ decays, while the
heavy quark spin symmetry relates semileptonic $b \to u$ decays
to the rare $b \to s$ decays.

{}From the point of view of heavy quark symmetries the simplest objects
are the $\Lambda$-type baryons, in which the light degrees of freedom
form a spin-zero state.

It has been observed that for the case of such a heavy baryon
decaying into a light spin 1/2 baryon, heavy quark symmetries
indeed restrict
the number of independent form factors \cite{MRRhtol}. To leading
order in the
$\bar\Lambda /m_Q$ expansion, this type of transition is described
by only two form factors. In addition, these two form factors enter
in such a combination that the ratio $G_A / G_V$ turns out to be
minus one. This fact has also interesting consequences for the
heavy mass expansion in the decay
$\Lambda_b \to \Lambda_c \ell \nu$ \cite{MR92}.

This prediction, namely $G_A / G_V = -1$, has been confirmed
by experiments. The first indirect measurement of this parameter has
been performed in a nonleptonic decay, where the polarization variable
$\alpha$\footnote{For the definition of the polarization variables
                  in baryon nonleptonic decays see \cite{PDG}.}
in the nonleptonic decay $\Lambda_c \to \Lambda \pi$ was measured.
Both ARGUS \cite{ARGUSnl} and CLEO \cite{CLEOnl} observe polarizations
which are consistent with
$\alpha = -1$. If factorization is assumed to relate the decay
$\Lambda_c \to \Lambda \pi$ to semileptonic $\Lambda_c \to \Lambda$
transitions the observed
value exactly corresponds to $G_A / G_V = -1$ predicted by heavy quark
symmetry. However, this is a nonleptonic decay and factorization has
been assumed ad hoc; hence this result may be as well accidental.

A more stringent test of this prediction requires the measurement
of the $\Lambda_c \to \Lambda $ semileptonic decay mode. Data
on this has become available recently;
both ARGUS \cite{ARGUSsl} and CLEO \cite{CLEOsl} performed
a measurement of the
polarization in the semileptonic decay $\Lambda_c \to \Lambda \ell \nu$
with results consistent with the heavy quark symmetry prediction.
The errors on both measurements are still at a level of 30 percent
and are thus quite large; however, these errors will be reduced in the
near
future.

As data becomes more precise,
one has to include corrections to the
heavy quark limit prediction using heavy quark effective field theory.
The leading logarithmic QCD correction will not modify the result for
$G_A / G_V$ since both the axial and the vector current scale in the
same way at scales below the heavy quark mass.
Consequently, the leading corrections will be the ones of order
$\Lambda_{QCD}/m_c$ which may be parametrized in terms of new form
factors. To some extend this analysis has been performed in
\cite{Ro92}; however, not all the form factors allowed by heavy
quark symmetries were considered in \cite{Ro92}.
In the present paper we shall investigate the subleading corrections to
the leading order result $G_A / G_V = -1$ using the heavy quark
effective theory. This amounts to expressing $G_A /G_V$ in terms of all
the leading and subleading
form factors in the $1/m_c$ expansion.  It turns out that this
expression involves too many form factors such that
an
estimate of the subleading effects requires additional input beyond the
heavy
quark effective theory. In the present paper we
use the limit of a heavy $s$ quark as a model assumption.

In the next section we shall give parametrizations for the leading
and the subleading form factors for a general baryonic heavy to light
transition
and count the number of independent form factors. In section 3 we
focus on transitions induced by the left handed current and extract
$G_A / G_V$ in terms
of the leading and subleading form factors.
Finally, we estimate the deviation to the lowest order relation $G_A /
G_V=-1$
arising from terms of the order $1/m_c$.

\section{Form Factors for Baryonic Heavy to Light Transitions}
It has been pointed out in \cite{Baryons,MRRhtol} that to leading
order
in the $1/m_c$ expansion only two independent form factors
$\Phi_1$ and $\Phi_2$ are needed to parametrize
the decay of a $\Lambda_c$ into a spin 1/2 light baryon
denoted here generically as $\Lambda$
\begin{equation}
< \Lambda (p) | \bar{q} \Gamma c_v | \Lambda_c (v) > =
\bar{u} (p) \left[ \Phi_1 + \Phi_2 \slash{v} \right] \Gamma u(v)
\end{equation}
Here $q$ is some light quark, $c_v$ is the static heavy $c$ quark
and $\Gamma$ is an arbitrary collection of Dirac matrices.

For the case of a left handed current $\Gamma = \gamma_\mu
(1-\gamma_5)$,
the above transition can be parametrized as follows:
\begin{eqnarray}
< \Lambda (p) | \bar{q} \gamma_\mu (1-\gamma_5) c | \Lambda_c (v) > &=&
\bar{u} (p) \left[ f_1 \gamma_\mu
               + i f_2 \sigma_{\mu \nu} q^\nu
               +   f_3 q^\mu  \right] u(p')  \nonumber \\
&+& \bar{u} (p) \left[ g_1 \gamma_\mu
               + i g_2 \sigma_{\mu \nu} q^\nu
               +   g_3 q^\mu  \right] \gamma_5 u(p')
\end{eqnarray}
where $p' = m_{\Lambda_c} v$ is the momentum of the $\Lambda_c$ whereas
$q = m_{\Lambda_c} v - p$ is the momentum transfer. From this one
defines the ratio $G_A / G_V$ by
\begin{equation}
\frac{G_A}{G_V} = \frac{g_1 (q^2 = 0)}{f_1 (q^2 = 0)}
\end{equation}

In the heavy $c$ quark limit one may relate the six form factors
$f_i$ and $g_i$ ($i=1,2,3$) to the two form factors $\Phi_j$
($j=1,2$)
\begin{eqnarray}
f_1 &=& - g_1 = \Phi_1 + \frac{m_\Lambda}{m_{\Lambda_c}} \Phi_2 \\
f_2 &=& f_3 = -g_2 = -g_3 = \frac{1}{m_{\Lambda_c}} \Phi_2
\end{eqnarray}
from which one reads off $G_A / G_V = -1$.

The $O(1/m_c)$-contributions to $G_A / G_V$ originate from
two sources. From the matching of the field operator of the $c$ quark
one has, to the order $1/m_c$, the replacement
\begin{equation}
c \longrightarrow c_v + \frac{1}{2 m_c} i \slash{D} c_v
\end{equation}
where $D$ is the covariant derivative of QCD acting on the $c$ quark.
This replacement will introduce local correction terms of the form
\begin{equation}
\Delta_l =  \frac{1}{2 m_c}
< \Lambda (p) |
         \bar{q} \Gamma i \slash{D} c_v | \Lambda_c (v) >
\end{equation}
In addition to that, one also obtains contributions from the correction
of the Lagrangian to order $1/m_c$ which are given in terms of three
operators
\begin{eqnarray} \label{l1}
&& {\cal L}_1 =  \bar{c}_v \frac{(i v \cdot D)^2}{2 m_c} h_c
\quad   \quad \quad
{\cal L}_2 =  \bar{c}_v \frac{(i D)^2}{2 m_c} c_v
\\ \nonumber
&& {\cal L}_3 = -\frac{g}{4 m_c} \bar{c}_v
             \sigma_{\alpha \beta} G^{\alpha \beta} c_v
\end{eqnarray}
These operators yield nonlocal terms of the form
\begin{equation}
\Delta_{nl} = i \int d^4 x \, \sum_{j=1}^3
< \Lambda (p) | T\left[ {\cal L}_j (x)
                \bar{q} (0) \Gamma c_v (0) \right] | \Lambda_c (v) >
\end{equation}
These relations are obtained by matching at the scale of the $c$ quark;
the renormalization group flow from scaling down to smaller scales will
introduce mixing of these operators with other operators of the
same dimension \cite{FG91,EH90}.

We start the discussion with the local terms and consider the matrix
element
\begin{equation}
R_\mu = < \Lambda (p) |
\bar{q} \Gamma i D_\mu  c_v | \Lambda_c (v) >
\end{equation}
which may be rewritten due to spin symmetry as
\begin{equation}
R_\mu = \bar\Lambda \bar{u} (p) {\cal M}_\mu \Gamma u(v)
\end{equation}
The scale of $R$ is set by the parameter $\bar\Lambda$ defined by
\begin{equation}
\bar\Lambda = m_{\Lambda_c} - m_c = {\cal O} (\Lambda_{QCD}) ;
\end{equation}
we have made the dependence on $\bar\Lambda$ explicit
in order to obtain dimensionless quantities of order one in what
follows.
The Dirac matrix ${\cal M}_\mu$ describes the light degrees of freedom
and may be expanded in terms of the sixteen independent Dirac matrices.
It is constrained by the equation of motion of the heavy quark
\begin{equation}
v\cdot R = < \Lambda (p) | \bar{q} \Gamma (i v\cdot D)  c_v | \Lambda_c
(v) >
   = 0
\end{equation}
which yields $v\cdot {\cal M} = 0$. To this end, ${\cal M}_\mu$ may
thus
be expressed in terms of four scalar functions $A,B,C$ and $D$
\begin{equation}
{\cal M}_\mu = (A+\slash{v}B) \left(\frac{p_\mu}{v\cdot p} - v_\mu
\right)
 + i C \sigma_{\mu \nu} v^\nu
 + i D \varepsilon_{\mu \alpha \beta \rho} v^\alpha
\frac{p^\beta}{v\cdot p}
        \gamma^\rho \gamma_5
\end{equation}
where $A,B,C$ and $D$ depend on the variable $v\cdot p$ and do not
scale
with the heavy mass for a fixed $v\cdot p$.

In fact these four form factors are not entirely independent from  the
leading
ones. First of all, momentum conservation implies
\begin{eqnarray} \label{mc0}
i \partial_\mu
< \Lambda (p) | \bar{q} \Gamma  c_v | \Lambda_c (v) > &=&
- (p_\mu - \bar{\Lambda} v_\mu )
< \Lambda (p) | \bar{q} \Gamma  c_v | \Lambda_c (v) > \\
= < \Lambda (p) | \overline{-iD_\mu q} \Gamma  c_v | \Lambda_c (v)  >
&+& \nonumber
< \Lambda (p) | \bar{q} \Gamma  iD_\mu c_v | \Lambda_c (v) >
\end{eqnarray}
This relation may be exploited by inserting
$\Gamma \to \gamma^\mu \Gamma$ and by contracting the index $\mu$.
Using the equation of motion for the light quark $i\slash{D}q = m_q q$
one obtains
\begin{equation} \label{mc}
\bar{u} (p) \left[ \Phi_1 + \slash{v} \Phi_2 \right]
\left( \bar{\Lambda} \slash{v} - \slash{p} + m_q \right) \Gamma u(v)
= \bar\Lambda \bar{u} (p) {\cal M}_\mu \gamma^\mu \Gamma u(v)
\end{equation}
This equation leads to two relations between the leading and the
subleading form factors
\begin{eqnarray} \label{momcon}
\left(\bar{\Lambda} - 2 v\cdot p \right) \Phi_2 - ( m_\Lambda - m_q )
\Phi_1
&=&
\bar\Lambda \left( \frac{m_\Lambda}{v\cdot p} A + B - 2D \right) \\
\nonumber
\bar{\Lambda} \Phi_1 + ( m_\Lambda + m_q ) \Phi_2
&=&
\bar\Lambda \left( -A - \frac{m_\Lambda}{v\cdot p} \left(B - 2D \right)
+ 3 C \right)
\end{eqnarray}
The second equation allows to express $C$ in terms of $A$ alone,
while the first equation may be used to express D in terms of $A$ and
$B$. We shall treat the four form factors $\Phi_1, \Phi_2,A$ and $B$
as independent quantities.

The nonlocal contributions will introduce more independent form
factors. Due to the equation of motion of the heavy quark, the first
term ${\cal L}_1$ will not contribute.
The second term ${\cal L}_2$ is a
singlet under spin symmetry and hence will yield a
renormalization of the leading order form factors
\begin{eqnarray}
&& i \int d^4 x \,
< \Lambda (p) | T\left[ ({\cal L}_1 (x) + {\cal L}_2 (x) )
                \bar{q} (0) \Gamma c_v (0) \right] | \Lambda_c (v) >
\\ \nonumber
&& = \frac{\bar\Lambda}{2m_c} \bar{u} (p)
\left[ \delta_1 + \delta_2 \slash{v} \right] \Gamma u(v) .
\end{eqnarray}

Finally, the magnetic moment operator ${\cal L}_3$ may be written as
\begin{equation}
i \int d^4 x \,
< \Lambda (p) | T\left[ {\cal L}_3 (x)
                \bar{q} (0) \Gamma c_v (0) \right] | \Lambda_c (v) >
= i \frac{\bar\Lambda}{2m_c} \bar{u} (p) T^{\alpha \beta}
  \Gamma \frac{\slash{v}+1}{2} \sigma_{\alpha \beta} u(v)
\end{equation}
where the Dirac matrix $T_{\mu \nu}$ introduces another five
independent form factors
\begin{equation}
T_{\mu \nu} = -i\rho_1 \sigma_{\mu\nu}
            + \rho_2 ( p_\mu \gamma_\nu-p_\nu \gamma_\mu )
         + i \varepsilon_{\mu \nu \alpha \beta} v^\alpha p^\beta
              (\rho_3 + \slash{v} \rho_4) \gamma_5
         + i \varepsilon_{\mu \nu \alpha \beta} v^\alpha \gamma^\beta
             \gamma_5 \rho_5
\end{equation}

In the following we shall concentrate on the discussion of the left
handed current and calculate the form factors $f_i$ and $g_i$ in terms
the ones parametrizing the leading and subleading contributions.

\section{Subleading Terms of the Left Handed Current}
For the left handed current one simply takes
$\Gamma =\gamma_\mu (1-\gamma_5)$
and re-arrange the subleading contributions
according to the parametrization shown in Eq. (2).

The nonlocal contribution involving ${\cal L}_2$ is a singlet under
spin symmetry; hence the two form factors $\delta_1$ and $\delta_2$
only renormalize the leading order form factors $\Phi_1$ and
$\Phi_2$.
\begin{equation}
\tilde{\Phi}_1 = \Phi_1+ \delta_1 \frac{\bar\Lambda}{2m_c}, \qquad
\tilde{\Phi}_2 = \Phi_2+ \delta_2 \frac{\bar\Lambda}{2m_c}.
\end{equation}
This result is in fact independent of $\Gamma$.

Furthermore, the nonlocal contribution coming from the
chromomagnetic moment operator ${\cal L}_2$ has in total five
form factors. However, two of them, $\rho_3$ and $\rho_4$ only
renormalize two of the form factors parametrizing the local $1/m_c$
contributions
\begin{equation}
\tilde{A} = A+2v\cdot p\rho_3 , \qquad
\tilde{B} = B+2v\cdot p\rho_4.
\end{equation}
 In contrast to the previous case, this relation holds only in the case of a
left-handed current.

In terms of these redefined form factors one obtains for the
form factors $f_i$ and $g_i$ the following result
\begin{eqnarray}
f_1 &=& \tilde{\Phi}_1 + \frac{m_\Lambda}{m_{\Lambda_c}} \tilde{\Phi}_2
\\
&& + \frac{\bar\Lambda}{2m_c}
     \left[ - \tilde{A} (1-\frac{m_{\Lambda_c}}{v \cdot p})
       + \tilde{B} \left(\frac{m_\Lambda}{m_{\Lambda_c}}
                   (\frac{m_\Lambda}{v \cdot p} - 1) - 2
                + \frac{m_\Lambda + m_{\Lambda_c}}{v \cdot p} \right)
      \right. \nonumber \\
&& - C (1+2\frac{m_\Lambda}{m_{\Lambda_c}})
- \frac{D}{v \cdot p}(m_{\Lambda_c}-\frac{m^2_\Lambda}{m_{\Lambda_c}})
+ \rho_1(2+4\frac{m_\Lambda}{m_{\Lambda_c}}) \nonumber \\
&& \left. - 2\rho_2(m_{\Lambda_c}-\frac{m^2_\Lambda}{m_{\Lambda_c}})
- 2\rho_5(2+\frac{m_\Lambda}{m_{\Lambda_c}}) \right] \nonumber \\
&& \qquad \nonumber \\
f_2 &=& \frac{1}{m_{\Lambda_c}} \tilde{\Phi}_2 \\
&& + \frac{\bar\Lambda}{2m_c}
     \left[ \tilde{A} \frac{1}{v \cdot p}
       + \tilde{B} \left(\frac{1}{v \cdot p}
                   (\frac{m_\Lambda}{m_{\Lambda_c}} + 1)
                - \frac{1}{m_{\Lambda_c}}\right)
      \right. \nonumber \\
&& - 2C \frac{1}{m_{\Lambda_c}}
+ \frac{D}{v \cdot p}(\frac{m_\Lambda}{m_{\Lambda_c}}-1)
+ 4 \rho_1 \frac{1}{m_{\Lambda_c}} \nonumber \\
&& \left. - 2\rho_2(1-\frac{m_\Lambda}{m_{\Lambda_c}})
- 2\rho_5\frac{1}{m_{\Lambda_c}} \right] \nonumber \\
&& \qquad \nonumber \\
f_3 &=& \frac{1}{m_{\Lambda_c}} \tilde{\Phi}_2 \\
&& + \frac{\bar\Lambda}{2m_c}
     \left[ -\tilde{A} \frac{1}{v \cdot p}
       + \tilde{B} \left(\frac{1}{v \cdot p}
                   (\frac{m_\Lambda}{m_{\Lambda_c}} - 1)
                - \frac{1}{m_{\Lambda_c}}\right)
      \right. \nonumber \\
&& - 2C \frac{1}{m_{\Lambda_c}}
+ \frac{D}{v \cdot p}(\frac{m_\Lambda}{m_{\Lambda_c}}+1)
+ 4 \rho_1 \frac{1}{m_{\Lambda_c}} \nonumber \\
&& \left. + 2\rho_2(1+\frac{m_\Lambda}{m_{\Lambda_c}})
- 2\rho_5\frac{1}{m_{\Lambda_c}} \right] \nonumber \\
&& \qquad \nonumber \\
g_1 &=& -\tilde{\Phi}_1 - \frac{m_\Lambda}{m_{\Lambda_c}}\tilde{\Phi}_2
\\
&& + \frac{\bar\Lambda}{2m_c}
     \left[ \tilde{A} (1-\frac{m_{\Lambda_c}}{v \cdot p})
       + \tilde{B} \left(\frac{m_\Lambda}{m_{\Lambda_c}}
                   (\frac{m_\Lambda}{v \cdot p} + 1) - 2
                - \frac{m_\Lambda - m_{\Lambda_c}}{v \cdot p} \right)
      \right. \nonumber \\
&& + C (1-2\frac{m_\Lambda}{m_{\Lambda_c}})
- \frac{D}{v \cdot p}(m_{\Lambda_c}-\frac{m^2_\Lambda}{m_{\Lambda_c}})
- \rho_1(2-4\frac{m_\Lambda}{m_{\Lambda_c}}) \nonumber \\
&& \left. - 2\rho_2(m_{\Lambda_c}-\frac{m^2_\Lambda}{m_{\Lambda_c}})
- 2\rho_5(2-\frac{m_\Lambda}{m_{\Lambda_c}}) \right] \nonumber \\
&& \qquad \nonumber \\
g_2 &=& -\frac{1}{m_{\Lambda_c}} \tilde{\Phi}_2 \\
&& + \frac{\bar\Lambda}{2m_c}
     \left[ \tilde{A} \frac{1}{v \cdot p}
       + \tilde{B} \left(\frac{1}{v \cdot p}
                   (\frac{m_\Lambda}{m_{\Lambda_c}} - 1)
                + \frac{1}{m_{\Lambda_c}}\right)
      \right. \nonumber \\
&& - 2C \frac{1}{m_{\Lambda_c}}
+ \frac{D}{v \cdot p}(\frac{m_\Lambda}{m_{\Lambda_c}}+1)
+ 4 \rho_1 \frac{1}{m_{\Lambda_c}} \nonumber \\
&& \left. + 2\rho_2(1+\frac{m_\Lambda}{m_{\Lambda_c}})
+ 2\rho_5\frac{1}{m_{\Lambda_c}} \right] \nonumber \\
&& \qquad \nonumber \\
g_3 &=& - \frac{1}{m_{\Lambda_c}} \tilde{\Phi}_2 \\
&& + \frac{\bar\Lambda}{2m_c}
     \left[ -\tilde{A} \frac{1}{v \cdot p}
       + \tilde{B} \left(\frac{1}{v \cdot p}
                   (\frac{m_\Lambda}{m_{\Lambda_c}} + 1)
                + \frac{1}{m_{\Lambda_c}}\right)
      \right. \nonumber \\
&& - 2C \frac{1}{m_{\Lambda_c}}
+ \frac{D}{v \cdot p}(\frac{m_\Lambda}{m_{\Lambda_c}}-1)
+ 4 \rho_1 \frac{1}{m_{\Lambda_c}} \nonumber \\
&& \left. - 2\rho_2(1-\frac{m_\Lambda}{m_{\Lambda_c}})
+ 2\rho_5\frac{1}{m_{\Lambda_c}} \right] \nonumber
\end{eqnarray}

These expressions are quite lengthy and one has to point out
that after redefinitions of $\Phi_1$, $\Phi_2$, $A$ and $B$ there are
in total nine independent form factors parametrizing the
leading and subleading contributions.  Note that one can no longer apply
(\ref{momcon})
to eliminate $C$ and $D$ since one has already redefined $\Phi_1$, $\Phi_2$,
etc.  It is well known that the most general parametrization of
the left handed current needs only six form factors; hence the heavy quark
limit alone will not be useful once the subleading corrections are
taken into account. In order to proceed further one needs additional
input beyond heavy quark symmetry which will be discussed in the
next section.

\section{Discussion of the Result}
Some insight into the anatomy of the subleading corrections
may be obtained by considering formally the
infinite mass limit also for the quark $q$. In that case all the
form factors which violate the spin symmetry of the quark $q$
vanish in the leading order. This implies that the form factor $\Phi_2$
should be excluded from the leading-order structure.  For local
$1/m_c$-corrections, only the form factor $A$ is allowed
by spin symmetry; all other form factors are suppressed by $1/m_q$.
In addition, it has been shown in \cite{GGW91} that the
subleading from factor $A$ may be related to the leading one $\Phi_1$.
In the present formalism this result may be obtained
by observing that in the heavy mass limit for the quark $q$, one
has, to the leading order in the $1/m_q$ expansion, $m_{\Lambda_c} -
m_Q =
m_\Lambda - m_q = \bar\Lambda$. Furthermore, the spin symmetry of $q$,
(\ref{mc0}) implies the substitution
$\Gamma \to \frac{1+\slash{v}'}{2}\Gamma$, where $\slash{v}'$ is the four
velocity of $q$. Due to the projection $\frac{1+\slash{v}'}{2}$
$\slash{v}$ in (\ref{mc}) becomes replaced by $v\cdot v'$ and the
two relations of (\ref{momcon}) collapse into a single one which is

\begin{equation}
\Phi_1 (v\cdot v' -1) = \frac{A}{v\cdot v'} (1-(v\cdot v')^2) .
\end{equation}
This exactly reproduces the result obtained in \cite{GGW91}.

However, as it was shown in \cite{MR92}, the linear terms in both
$1/m_c$ as well as $1/m_q$ do not change the relation $f_1 = - g_1$
and thus one still has $G_A / G_V = -1$ to this order. Terms, which
change this relation, will thus be of the order of
$\bar\Lambda^2/(m_c m_q)$ in a framework, where an expansion in inverse
powers of the mass $m_q$ makes sense. Terms of this kind have been
studied in \cite{FN92}; however, for a quantitative estimate of these
terms, an input beyond the heavy quark effective theory is necessary.

In \cite{FN92} the ratio $G_A / G_V$ is studied for the case
of $\Lambda_b \to \Lambda_c e \nu$. In this case the ratio $G_A / G_V$
depends on three unknown parameters which are ratios of form
factors, taken at the point $q^2 = 0$. The corrections in this case
have been estimated to be of the order of a few percent, thus being
of the typical order of magnitude expected from the quantity
$\bar\Lambda^2/(m_c m_b) \sim 1\%$.
It has been pointed out \cite{FN92} that the quantity which is
accessible
experimentally is the polarization variable $\alpha_{\Lambda_c}$.
The relation of this variable to the ratio of form factors is given
by
\begin{equation}
\alpha_{\Lambda_c} = \frac{2x}{1+x^2} \mbox{ where }
x= G_A / G_V
\end{equation}
This variable is quite insensitive to corrections to $G_A / G_V$
if $G_A / G_V$ is close to $-1$: If $x=-1+\epsilon$,
then $\alpha_{\Lambda_c}$ receives only corrections of the order
$\epsilon^2$
\begin{equation}
\alpha_{\Lambda_c} = -1+\frac{1}{2} \epsilon^2
\end{equation}
This means that the polarization variable only receives corrections
of the order $1/m_c^2$, although the form factors and their ratios
will have corrections linear in $1/m_c$.

Another very rough estimate on the corrections
to the polarization variable may be obtained by using the
information of the
exclusive semileptonic $\Lambda$ decays. From the heavy quark
arguments
displayed above one would expect that $\alpha_\Lambda + 1$ scales
like $\bar\Lambda^2 / m_s^2$ in a limit, where the $s$
quark is heavy. In order to obtain an estimate for $\alpha_{\Lambda_c}$
we shall take this limit as a model assumption; we may then use the
measured value of $\alpha_\Lambda$ in the decay $\Lambda \to p
e \nu$ to scale up to the value of $\alpha_{\Lambda_c}$
\begin{equation}
0.282 \pm 0.015 = (\alpha_\Lambda + 1) = \frac{m_c^2}{m_s^2}
(\alpha_{\Lambda_c} + 1)
\end{equation}
where the measured value has been taken from \cite{PDG}. Taking
a value of 1.8 GeV for the $c$ quark mass and varying the mass of
the $s$ quark in the range between 400 and 600 MeV one obtains
for the corrections to the polarization variable
\begin{equation}
0.01 \le (\alpha_{\Lambda_c} + 1) \le 0.04
\end{equation}
and hence one expects a fairly small deviation from unity.
However, the smallness of the correction is mainly due to the
fact that the polarization variable $\alpha$ is insensitive to
corrections to $G_A / G_V = -1$. If one translates this estimate
for $\alpha_{\Lambda_c}$ into a result for $G_A / G_V $ one has

\begin{equation}
0.14\le \vert G_A / G_V + 1 \vert \le 0.25
\end{equation}
This result is consistent with the estimate given for
$\Lambda_b \to \Lambda_c e \nu$ in \cite{FN92} if one scales
up the result there by a factor of $m_b / m_s \sim 10$. We have
to stress again that this is not meant to be a detailed
quantitative analysis, however, the qualitative agreement
with the scaled up result of \cite{FN92} and the result obtained
from the semileptonic $\Lambda$ decays give
a consistent picture of the corrections to be expected.

Finally we want to comment on perturbative QCD corrections.
They have been calculated and may be found in the literature
\cite{FG91,EH90}. The leading order result at the level of the leading
logarithmic approximation (LLA) is independent
of the Dirac structure of the current as a consequence of spin
symmetry \cite{HQET}; hence $G_A / G_V $ is not changed in LLA.

To order $1/m_c$, perturbative QCD introduces
additional operators; the matrix elements of these may be
related to the form factors defined above and hence these
corrections will not introduce additional unknown functions.
However, we have not given these QCD corrections here, since
 they will be small, of the order $\alpha_s (m_c)$ and
$(m_s / m_c) \alpha_s (m_c) \ln (m_c^2 / m_s^2)$, and a
quantitative estimate of the effects is difficult anyway.

\section{Conclusion}
In the present paper we have given a detailed discussion on
the leading recoil corrections for the case of
a heavy $\Lambda_c$ decaying into a light baryon. The prototype  of
such a transition is the decay $\Lambda_c \to \Lambda e \nu$
which has recently been measured by both ARGUS and CLEO.

The main prediction of heavy quark effective theory for this
class of decays is $G_A / G_V = -1$ which has
been checked in the recent experiments. It turns out that the
prediction indeed holds, although the errors are still quite
large.

Subleading corrections of the order of $\bar\Lambda / m_c$ will
change this result; however, it turns out to be difficult to
estimate this change quantitatively, since in total nine unknown
form factor enter the game at the order $\bar\Lambda / m_c$
and thus not much can be said from the heavy quark effective theory
alone.

A rough estimate may be obtained from the assumption that the
$s$ quark is heavy; however, we want to stress that this has
the character of a model assumption. We expect this to at least
yield the correct magnitude of the subleading terms, since
the assumption of a heavy $s$ quark has in fact yield reasonable
results, once the leading recoil corrections had been taken into
account \cite{heavy-s}.

{}From these estimates we obtain a correction between 20\% and 30\%
for the ratio $G_A / G_V$ which is of the expected size. However,
the experimentally accessible quantity is the polarization variable
$\alpha_{\Lambda_c}$ which depends only weakly on the corrections
to $G_A / G_V$. For this variable it is very likely that its
value is smaller than $-0.95$. Since $\alpha_{\Lambda_c}$ cannot
exceed $-1$, a check of this requires a measurement of
$\alpha_{\Lambda_c}$ to the precision of a few per cent. However, this
will require a reduction of the error of the present experimental
data by a factor of ten, which will be difficult to achieve.

\section*{Acknowledgements}
T.M. acknowledges clarifying discussions with M. Neubert and
D. Cassel.
G.L.L. appreciate the hospitality of the Institut f\"ur
Kernphysik, Darmstadt (Germany) where a part of this work
was done. G.L.L. is supported in part by the National Science Council
of the Republic of China, under the contract number
NSC83-0208-M-001-014.

\end{document}